\documentclass[aps,prl,twocolumn,amsmath,amssymb,superscriptaddress,citeautoscript,longbibliography,floatfix]{revtex4-1}
\newcommand{\angstrom}{\textup{\AA}}
\usepackage[allow-number-unit-breaks]{siunitx}
\usepackage{graphicx}
\usepackage{dcolumn}
\usepackage{bm}
\usepackage{color}
\usepackage{bbding}
\usepackage{pifont}

\usepackage{mathrsfs}
\usepackage{amsfonts}
\usepackage{multirow}
\usepackage{natbib}
\usepackage{epsf}
\usepackage{epsfig}
\usepackage{epstopdf}
\usepackage{amsmath}
\usepackage{hyperref}
\hypersetup{
    colorlinks=true,
    linkcolor=blue,
    filecolor=cyan,      
    citecolor=blue,
    urlcolor=blue,
    }

\begin{document}

\title{Reversing N\'eel Vector in $\mathcal{PT}$-Antiferromagnets by Nonreciprocal Light Scattering}

\author{Qianqian Xue}
\affiliation{Center for Alloy Innovation and Design, State Key Laboratory for Mechanical Behavior of Materials, Xi'an Jiaotong University, Xi'an, 710049, China}

\author{Jian Zhou}\email{jianzhou@xjtu.edu.cn}
\affiliation{Center for Alloy Innovation and Design, State Key Laboratory for Mechanical Behavior of Materials, Xi'an Jiaotong University, Xi'an, 710049, China}

\begin{abstract}
Antiferromagnetic (AFM) spintronics has been receiving tremendous attention due to their ultrafast kinetics, zero stray field, immune to external magnetic field, and potential to minimizing magnetic storage devices. The optical control of AFM N\'eel vector has become a hectic topic during recent years, which could facilitate the AFM utilization in practical systems. In this work, we propose a nonreciprocal light scattering mechanism to flip the N\'eel vector in parity-time ($\mathcal{PT}$) combined AFM multilayers, by estimating the energy contrast between the bistable N\'eel polarization configurations. We illustrate our theory using a low energy $\bm k\cdot\bm p$ model, and perform \textit{ab initio} calculations on two typical $A$-type AFM materials, $\mathrm{MnBi}_2\mathrm{Te}_4$ and $\mathrm{CrI}_3$ thin films. We show that varying incident photon frequency could modulate the relative stability between the bistable N\'eel vector state, which also depends on the light handedness. According to this theory, our parameter-independent calculations on the N\'eel vector diagram shows consistent predictive results with recent experimental observations. This mechanism provides an effective route to controlling the AFM order parameter through photo-magnetic interaction.
\end{abstract}

\maketitle

\textit{Introduction.} Manipulating different orders of parameter is one of the most essential aspects in condensed matter physics and materials science, leading to various pivotal practical applications. In addition to many ferroic (such as ferroelectric, ferromagnetic, and ferroelastic) orders that have been widely-used, recent years have witnessed increasing interests and extensive attention in antiferro-systems \cite{Kittel51PR,Keffer52PR,Jungwirth16NN,Baltz18RMP,Smejkal18NP,Han23NM}. For instance, antiferromagnetic (AFM) materials that possess zero intrinsic net magnetic moments have been demonstrated to exhibit ultrafast kinetics and high resolution spintronic processes, described by its order parameter such as the N\'eel vector $\mathcal{N}_0$ in collinear AFMs. The N\'eel vector direction determines the magnetic symmetry of the system (with spin-orbit coupling, SOC), hence it could precisely control and tune various transport and optical responses, such as anomalous Hall effect \cite{Chen14PRL,Nakatsuji15Nature,Kotegawa24PRL,Yang24PRB}, nonlinear Hall effect \cite{Gao23Science,Wang21PRL,Wang23PRL,Mei24PRB}, bulk photovoltaic effect \cite{Wang20npjCM,Zhang19NC,Fei20PRB,Zhang22NL,Pi23QuanFron,Wu24npjCM}, and valley polarization \cite{Li13PNAS,Xue23PRB}.

For practical applications of AFMs, a natural question is how to control and manipulate $\mathcal{N}_0$ \cite{Han24SA,Bodnar18NC,Reimers23NC,Godinho18NC,Kocsis18PRL}. As with vanishing magnetic moments, an magnetic field in general cannot change its direction and other recipes need exploration. Among them, optical approaches are very attractive as light is contactless and does not introduce unwanted impurities and disorders into the sample. The ultrafast nature also guarantees efficient control with high resolution. While many works focus on the light-induced magnetization reorientation and demagnetization in ferromagnets \cite{Zhang00PRL,Chovan06PRL,Henriques18PRL,Wu24NC}, it still remains an open question on how to understand the fundamental mechanism of light-induced $\mathcal{N}_0$ flip in compensated AFMs. Here, we focus on $A$-type AFM multilayers with $\mathcal{PT}$ (combined parity and time-reversal) symmetry, which usually show a bistable configuration with degenerate $-\mathcal{N}_0$ and $\mathcal{N}_0$, such as even layered $\mathrm{MnBi}_2\mathrm{Te}_4$ \cite{Deng20Science,Zhang19PRL,Otrokov19PRL} and $\mathrm{CrI}_3$ \cite{Huang17Nature,Zhang22NM} films. Both these materials have been experimentally demonstrated to exhibit an out-of-plane easy axis, and prefer intralayer ferromagnetic and interlayer AFM configuration (the former one shows an axion insulating feature). We propose a symmetry-adapted microscopic mechanism that incorporates nonreciprocal light scattering process as it normally irradiates on the multilayers. Accordingly, we derive a light-induced free energy variation formalism, to show how nonreciprocally propagating circularly polarized light (CPL) breaks $\mathcal{P}$ and $\mathcal{T}$ symmetries simultaneously. Hence, CPL lifts the energy degeneracy between $\mathcal{N}_0$ and $-\mathcal{N}_0$ states. We then adopt a simplified $\bm k\cdot\bm p$ model to illustrate this process, and perform first-principles density functional theory (DFT) to apply such a theory in both $\mathrm{MnBi}_2\mathrm{Te}_4$ and $\mathrm{CrI}_3$ bilayers. We suggest that depending on the CPL handedness ($\sigma\pm$) and light frequency (or wavelength), the relative stability between $\mathcal{N}_0$ and $-\mathcal{N}_0$ can be switched. Our parameter-free DFT calculation predictions on the wavelength-dependent relationship between $\sigma\pm$ and $\pm\mathcal{N}_0$ give consistent results with recent experimental measurements, validating our theoretical mechanism.

\begin{figure}[t]
    \centering
    \includegraphics[width=0.48\textwidth]{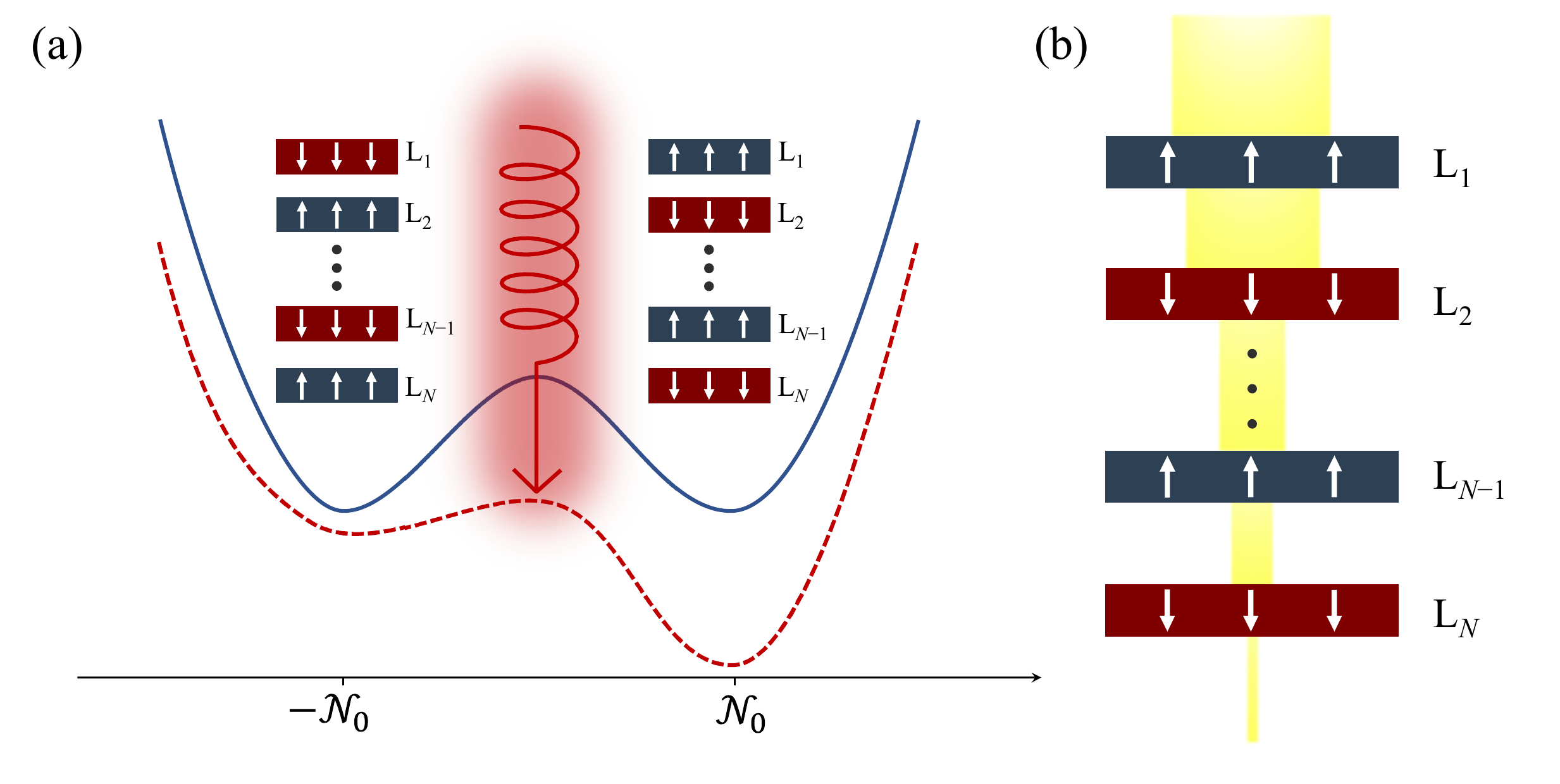}
    \caption{Schematic plot of circularly polarized light induced N\'eel flip. (a) Modulation of energy profile under light, with $\pm\mathcal{N}_0$ representing bistable states. (b) Light attenuation across layers. The scattering and attenuation break time-reversal $\mathcal{T}$ and inversion symmetry $\mathcal{P}$, respectively, and their synergistic effect could flip the N\'eel vector.}
    \label{fig:schematic}
\end{figure}

In the bistable $A$-type AFMs with parity-time $\mathcal{PT}$ symmetry, the $\mathcal{N}_0$ configuration is degenerate with $-\mathcal{N}_0$, protected by $\mathcal{P}$ and $\mathcal{T}$ [Fig. \ref{fig:schematic}(a)]. Hence, in order to lift their degeneracy and trigger N\'eel vector reversal, one has to simultaneously break both these two symmetries. We will focus on CPL irradiation here as it naturally breaks $\mathcal{T}$. When the sample thickness is on the order of a few to a few tens of nanometers (well-below light penetration depth) and laser pulses (rather than continuous wave) are assumed, the thermal effect can be safely eliminated. Hence, we will only consider athermic photomagnetic process \cite{Kirilyuk10RMP}.

\textit{Theory on light-induced free energy variation.} We consider normal incident light onto bistable AFM multilayers [Fig. \ref{fig:schematic}(b)]. The alternating magnetic field component of light is omitted as it is orders of magnitude smaller than the electric field components. Hence, the light is described by $\vec{\mathcal{E}}_{\omega}(t)=\mathrm{Re}\vec{E}e^{i\omega t}$. The in-plane electric field magnitude vector is $\vec{E}=E_{\omega}(1,\eta,0)$ with $\eta=\pm i$ indicating the right-handed $\sigma+$ and left-handed $\sigma-$ light. We use $\vec{e}^{\sigma\pm}=(1,\eta,0)$ to denote the electric field unit vector. Since the unit cell size is much smaller than the laser spot, the lateral inhomogeneity can be safely omitted ($\bm q=0$ interaction). Here, we assume that the energy transfer between light electric field and the sample stems from two aspects. Firstly, when the photon energy $\hbar\omega$ is above the sample bandgap $E_g$, light absorption naturally exists by exciting electron-hole pairs, and the internal energy gains. This is scaled by light absorbance. For ultrathin materials with thickness of $d$, the absorbance is
\begin{equation}\label{eq:abs}
    A^{\sigma\pm}_{\mathcal{N}_0}(\omega)=\frac{d}{c\varepsilon_0}\vec{e}^{{\sigma\pm},*}\cdot\tensor{\sigma}^{'}_{\mathcal{N}_0}(\omega)\cdot\vec{e}^{{\sigma\pm}},
\end{equation}
where $c$ denotes the speed of light in vacuum and $\varepsilon_0$ is vacuum permittivity. Here, $\tensor{\sigma}(\omega)$ refers to frequency-dependent optical conductivity, which relates to the susceptibility function by $\chi_{ij}(\omega)=\frac{i}{\varepsilon_0\omega}\sigma_{ij}(\omega)$. Eq. (\ref{eq:abs}) is valid when $d$ is much smaller than photon wavelength, measuring energy gain per area. Then the total free energy per area raises according to $\mathcal{G}^{\sigma\pm}_{\mathrm{abs},\mathcal{N}_0}(\omega)=nA\hbar\omega$ with $n$ describing expectation number of photons that irradiate on the sample per area.

On the other hand, we would like to introduce another mechanism that also affects the potential energy landscape. This is in analogous to the optical tweezer technique \cite{Pesce24OTbook} and has been largely overlooked previously. To the lowest order, it describes light scattering with respect to elementary excitations. Thermodynamically, the light scattering effect adds an additional potential energy (per area) in the form of \cite{Zhou21AS,Wu23JPCL}
\begin{equation}\label{eq:scatter}
    \mathcal{G}_{\mathrm{sc},\mathcal{N}_0}^{\sigma\pm}(\omega)=d\left[\vec{e}^{{\sigma\pm},*}\cdot\frac{\tensor{\sigma}^{''}_{\mathcal{N}_0}(\omega)}{4\omega}
    \cdot\vec{e}^{\sigma\pm}\right]E_{\omega}^2.
\end{equation}
Very recently, this light scattering effect has been used to predict nonvolatile phase transition in nonmagnetic materials under nonresonant light irradiation, and its validity has been quantitatively demonstrated via recent experiments \cite{Shi23NC,shi2024ArXiv}. In magnetic materials, broken of $\mathcal{T}$ yields finite nondiagonal components in $\tensor{\sigma}$. Recently, Eq. (\ref{eq:scatter}) is also derived according to nonequilibrium steady state theory of a fast oscillating system on a slow varying field, termed as ``ponderamotive'' potential \cite{Sun24PRB}. Hence, it belongs to Floquet engineering picture. According to Eq. (\ref{eq:scatter}), $\mathcal{G}_{\mathrm{sc}}(\omega)$ scales with $E_{\omega}^2$ (or light intensity $I$), hence it lies on the same order with $\mathcal{G}_{\mathrm{abs}}(\omega)$. In a simple damped harmonic model, one sees $\mathcal{G}_{\mathrm{sc}}\propto\mathrm{Re}[1/(\omega^2-\omega_0^2+i\gamma\omega)]$. Thus, it could occur when photon energy is above and below $E_g$. As a scattering process, it keeps the photon number $n$ to be unchanged.

Our light-induced N\'eel vector flipping mechanism is established upon the above two mechanisms. When the CPL irradiates onto the multilayer, light attenuation occurs as each van der Waals (vdW) layer absorbs the photon flux [Fig. \ref{fig:schematic}(b)]. This corresponds to a nonreciprocal process that breaks $\mathcal{P}$. In particular, when the normal incident CPL irradiates onto the multilayer (with a total $N$ vdW layers) along $-z$, at the $i^{\mathrm{th}}$ vdW layer, its free energy per area varies according to
\begin{equation}
\begin{split}
    &\mathcal{G}^{\sigma\pm}_{\mathcal{N}_0}(\mathrm{L}_i,\omega)=\mathcal{G}_{0,\mathcal{N}_0}(\mathrm{L}_i)+\mathcal{G}_{\mathrm{sc},\mathcal{N}_0}^{\sigma\pm}(\mathrm{L}_i,\omega)+\mathcal{G}_{\mathrm{abs},\mathcal{N}_0}^{\sigma\pm}(\mathrm{L}_i,\omega) \\
    &=\mathcal{G}_{0,\mathcal{N}_0}(\mathrm{L}_i)+d\lambda_{\mathcal{N}_0}^{\sigma\pm}(\mathrm{L}_i,\omega)\left[\vec{e}^{{\sigma\pm},*}\cdot\frac{\tensor{\sigma}^{''}_{\mathcal{N}_0}(\mathrm{L}_i,\omega)}{4\omega}
    \cdot\vec{e}^{\sigma\pm}\right]E_{\omega}^2 \\
    &+\frac{d}{c\varepsilon_0}\vec{e}^{{\sigma\pm},*}\cdot\tensor{\sigma}^{'}_{\mathcal{N}_0}(\mathrm{L}_i,\omega)\cdot\vec{e}^{{\sigma\pm}}n_i\hbar\omega.
\end{split}
\end{equation}
Here, $\mathcal{G}_{0,\mathcal{N}_0}(\mathrm{L}_i)$ denotes the free energy before CPL is irradiated. As the conductance and susceptibility tensor is scaled by $1/d$ (see below) in practical calculations, the specific choice of $d$ value does not change the results for total energy variation. For $A$-type AFM multilayers, it is natural that $\sum_{i=1}^N\mathcal{G}_{0,\mathcal{N}_0}(\mathrm{L}_i)=\sum_{i=1}^N\mathcal{G}_{0,-\mathcal{N}_0}(\mathrm{L}_i,)$, with $N$ being an even number denoting the total layer numbers. As light is attenuated, $n_i$ monotonously reduces from $n_0$ (total photon number per area in vacuum within the pulse) under above $E_g$ light illumination. Here, we introduce an attenuation factor $\lambda^{\sigma\pm}_{\mathcal{N}_0}(\mathrm{L}_i,\omega)=n_i/n_0=\prod_{j=1}^{i-1}[1-A^{\sigma\pm}_{\mathcal{N}_0}(\mathrm{L}_j,\omega)]$ that accounts for photon number elimination. We use the electromagnetic wave expectation energy flux to replace photon energy, and one has $n_i\hbar\omega=c\varepsilon_0\lambda^{\sigma\pm}_{\mathcal{N}_0}(\mathrm{L}_i,\omega)\tau E_{\omega}^2$, with $\tau$ being the light duration time (usually within one picosecond in a few pulses, taken to be $0.1\,\mathrm{ps}$ in our study). In this regard, the total energy per unit area profile is renormalized according to
\begin{equation}\label{eq:gfe}
    \mathcal{G}^{\sigma\pm}_{\mathcal{N}_0}(\omega)=\sum_{i=1}^N\left[\mathcal{G}_{0,\mathcal{N}_0}(\mathrm{L}_i)+(\vec{e}^{\sigma\pm,*}\cdot\tensor{K}\cdot\vec{e}^{\sigma\pm})E_{\omega}^2\right],
\end{equation}
in which we have
\begin{equation}\label{eq:kernal}
\begin{split}
    \tensor{K}_{\mathcal{N}_0}(\mathrm{L}_i,\omega)&=\lambda^{\sigma\pm}d\left(\frac{\tensor{\sigma}^{''}_{}}{4\omega}+\tau\tensor{\sigma}^{'}\right) \\
    &=\varepsilon_0\lambda^{\sigma\pm}d\left(-\frac{1}{4}\tensor{\chi}^{'}+\tau\omega\tensor{\chi}^{''}\right).
\end{split}
\end{equation}
In this equation, both optical conductance $\tensor{\sigma}$ and susceptibility $\tensor{\chi}$ depend on layer index $\mathrm{L}_i$, N\'eel vector $\mathcal{N}_0$, and incident frequency $\omega$, which are omitted for clarity reason. Therefore, we can estimate the energy difference between $\mathcal{N}_0$ and $-\mathcal{N}_0$ states according to $\Delta\mathcal{G}^{\sigma\pm}=\mathcal{G}^{\sigma\pm}_{-\mathcal{N}_0}(\omega)-\mathcal{G}^{\sigma\pm}_{\mathcal{N}_0}(\omega)=\vec{e}^{\sigma\pm,*}\cdot[\sum_{i=1}^N\tensor{K}_{-\mathcal{N}_0}(\mathrm{L}_i,\omega)-\tensor{K}_{\mathcal{N}_0}(\mathrm{L}_i,\omega)]\cdot\vec{e}^{\sigma\pm}E_{\omega}^2$, which scales with the incident light intensity. Its sign depends on the incident photon energy $\hbar\omega$. Note that the above expression is not limited in vdW multilayers, and we give a continuum model for thin films in Supplemental Material \cite{supp,Kresse93PRB,Kresse96PRB,Perdew96PRL,Perdew08PRL,Blochl94PRB,Monkhorst76PRB,Heyd03JCP,Heyd06JCP,Mostofi14CPC}. 

\textit{Low energy model calculation.}$-$ We illustrate the above theory using a minimum $\bm k\cdot\bm p$ model \cite{Lei20PNAS,Burkov11PRL}
\begin{equation}
    \begin{split}
        H(\bm k)=&\sum_{i,j=1}^N\left\{(-1)^i[\hbar v_D(\hat{\bm z}\times\bm\sigma)\cdot\bm k+m_i^z\sigma_z]\delta_{ij}\right. \\
        +&\left.\Delta(\delta_{j,i+1}+\delta_{j,i-1})/2\right\}c_{\bm k,i}^{\dagger}c_{\bm k,j}.
    \end{split}
\end{equation}
Here, $i$ (and $j$) denotes vdW layer index in the system, and $\bm\sigma$ is the Pauli matrix for spin degree of freedom. $v_D$ is effective velocity at the Fermi energy, $\Delta$ denotes the vdW interactions between adjacent layers, and $m_i^z=-m_{i\pm 1}^z$ is the alternative $z$-component magnetization of each layer.

We take a structure with six layers in this model, and the parameters are taken as $\Delta=0.5v_D$ and $m_z=\pm2v_D$ (for $\mathcal{N}_0$ and $-\mathcal{N}_0$ states). Their parameter dependent results are summarized in Figs. S1 and S2 \cite{supp}, suggesting that it is the $m_z$ largely controls the optical susceptibility. We then calculate their susceptibility function $\chi^{\sigma\pm}(\omega)$ components. According to the independent particle approximation \cite{PhysRevB.73.045112}, the layer-resolved susceptibility is
\begin{equation}
    \begin{split}
        \chi_{\alpha\beta}(\mathrm{L}_j,\omega)&=-\frac{e^2L_c}{\varepsilon_0Nd}\int\frac{d^3\bm k}{(2\pi)^3}\sum_{n,m}(f_{n\bm k}-f_{m\bm k}) \\
        &\times\frac{\langle u_{n\bm k}|\nabla_{k_{\alpha}}\hat{P}_j|u_{m\bm k}\rangle\langle u_{m\bm k}|\nabla_{k_{\beta}}|u_{n\bm k}\rangle}{\hbar(\omega_{n\bm k}-\omega_{m\bm k}-\omega-i/\tau_{\mathrm{el}})}.
    \end{split}
\end{equation}
Here, $|u_{n\bm k}\rangle$, $\hbar\omega_{n\bm k}$, and $f_{n\bm k}$ represent the Bloch wavefunction, eigenenergy, and Fermi-Dirac occupation of state $n$ at $\bm k$, respectively. $\tau_{\mathrm{el}}$ denotes the electron scattering lifetime. Note that it actually depends on the specific band index and momentum in each Bloch state, but its exact evaluation is computationally demanding and not straightforward even in a perfect crystal. The disorder and impurities are also phenomenologically incorporated into $\tau_{\mathrm{el}}$. Hence, we follow the conventional method and take a universal value in our discussion. We also note that its exact value does not affect the main results in this work. $\hat{P}_j=\sum_{l\in\mathrm{L}_j}|u_l\rangle\langle u_l|$ denotes the projector operator on layer-$j$.

\begin{figure}[t]
    \centering
    \includegraphics[width=0.48\textwidth]{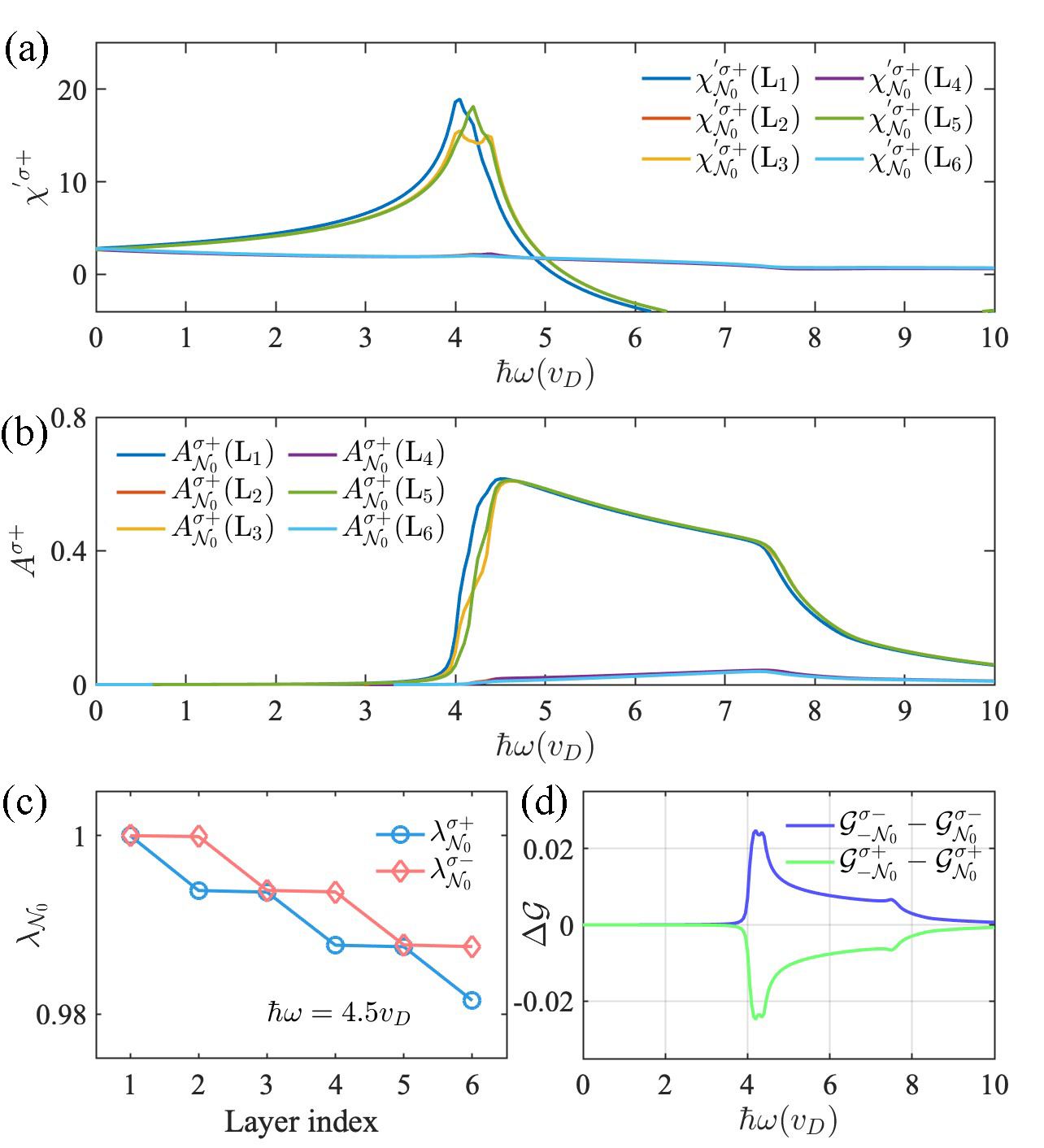}
    \caption{Low energy model results. Layer index dependent (a) real part and (b) absorbance spectrum of susceptibility function under right-handed CPL for a six layered system, in $\mathcal{N}_0$ state. The odd (even) index represents spin up (down) layer. (c) Attenuation factor $\lambda$ in the multilayer under left-handed and right-handed CPL. (d) Free energy difference between $\mathcal{N}_0$ and $-\mathcal{N}_0$ states, in arbitrary unit system.}
    \label{fig:kpmodel}
\end{figure}

The results are plotted in Fig. \ref{fig:kpmodel}. The real and imaginary parts of frequency dependent susceptibility in the whole system (summed over all six layers) satisfy the well-known Kramers-Kronig relation. The integral is performed in the 3D first Brillouin zone (BZ), and we introduce a scaling factor $\frac{L_c}{Nd}$ to eliminate the vacuum space contribution ($L_c$ is supercell lattice constant along $z$) \cite{Zhou21AS}. From Figs. \ref{fig:kpmodel}(a) and \ref{fig:kpmodel}(b), it is clearly seen that the $\sigma+$ light interacts with the spin up vdW layers stronger than the spin down layers in both $\chi^{'}(\omega)$ and $A(\omega)$, thus presenting an even-odd oscillation behavior. Under an inversion operation $\mathcal{P}$, one can find that the layer-resolved optical response function satisfies $\chi^{\sigma+,'}_{\mathcal{N}_0}(\mathrm{L}_i,\omega)=\chi^{\sigma+,'}_{-\mathcal{N}_0}(\mathrm{L}_{N+1-i},\omega)$ and $A^{\sigma+}_{\mathcal{N}_0}(\mathrm{L}_i,\omega)=A^{\sigma+}_{-\mathcal{N}_0}(\mathrm{L}_{N+1-i},\omega)$. Similarly, applying $\mathcal{T}$ one could yield $\sigma-$ illuminated results, $\chi^{\sigma-,'}_{\mathcal{N}_0}(\mathrm{L}_i,\omega)=\chi^{\sigma+,'}_{-\mathcal{N}_0}(\mathrm{L}_i,\omega)$ and $A^{\sigma-}_{\mathcal{N}_0}(\mathrm{L}_i,\omega)=A^{\sigma+}_{-\mathcal{N}_0}(\mathrm{L}_i,\omega)$. Our numerical results confirm these relationships. In Fig. \ref{fig:kpmodel}(c), we plot the layer dependent attenuation factor $\lambda^{\sigma\pm}(\mathrm{L}_i)$ at a typical frequency $\omega=4.5v_D$, slightly above the bandgap $E_g=4.0v_D$. One sees a step-wise monotonously reduction trend, depending on the CPL handedness and spin polarization of each layer. With the above results, we calculate the relative energy difference between the $\mathcal{N}_0$ and $-\mathcal{N}_0$ states under $\sigma\pm$ [Fig. \ref{fig:kpmodel}(d)]. When the photon energy is below $E_g$, even though the $\chi'$ would reduce the energy potential [Eq. (\ref{eq:scatter})], their energy difference remains to be zero as no light attenuation occurs. Above $E_g$, this model shows that $\sigma+$ ($\sigma-$) CPL favors $-\mathcal{N}_0$ ($\mathcal{N}_0$), under the synergistic effect of alternative light scattering and absorption over layers. Note that if the attenuation is ignored (setting $\lambda=1$), the two states always keep their energetic degeneracy. Similarly, without including the scattering process, their degeneracy also retains if only absorption exists.

We vary the total layer number to conduct the calculation. From Fig. S3 \cite{supp}, one sees that the bandgap of $A$-type AFM does not change significantly with respect to different layer numbers $N$. In addition, the sign and profile of $\Delta\mathcal{G}^{\sigma\pm}(\omega)$ remain to be independent of $N$. Interestingly, if $N$ is increased, to obtain the same value of $\Delta\mathcal{G}^{\sigma\pm}$, the laser intensity can be linearly reduced. This is due to stronger light scattering and absorption occur for larger $N$ system. Hence, in order to observe N\'eel vector flipping, one could conduct experimental measurement for relatively thicker multilayer systems. The thickness has to be constrained on the order of a few to a few tens of nanometers, below the light wavelength and penetration depth.

\begin{figure*}[t]
    \centering
    \includegraphics[width=0.9\textwidth]{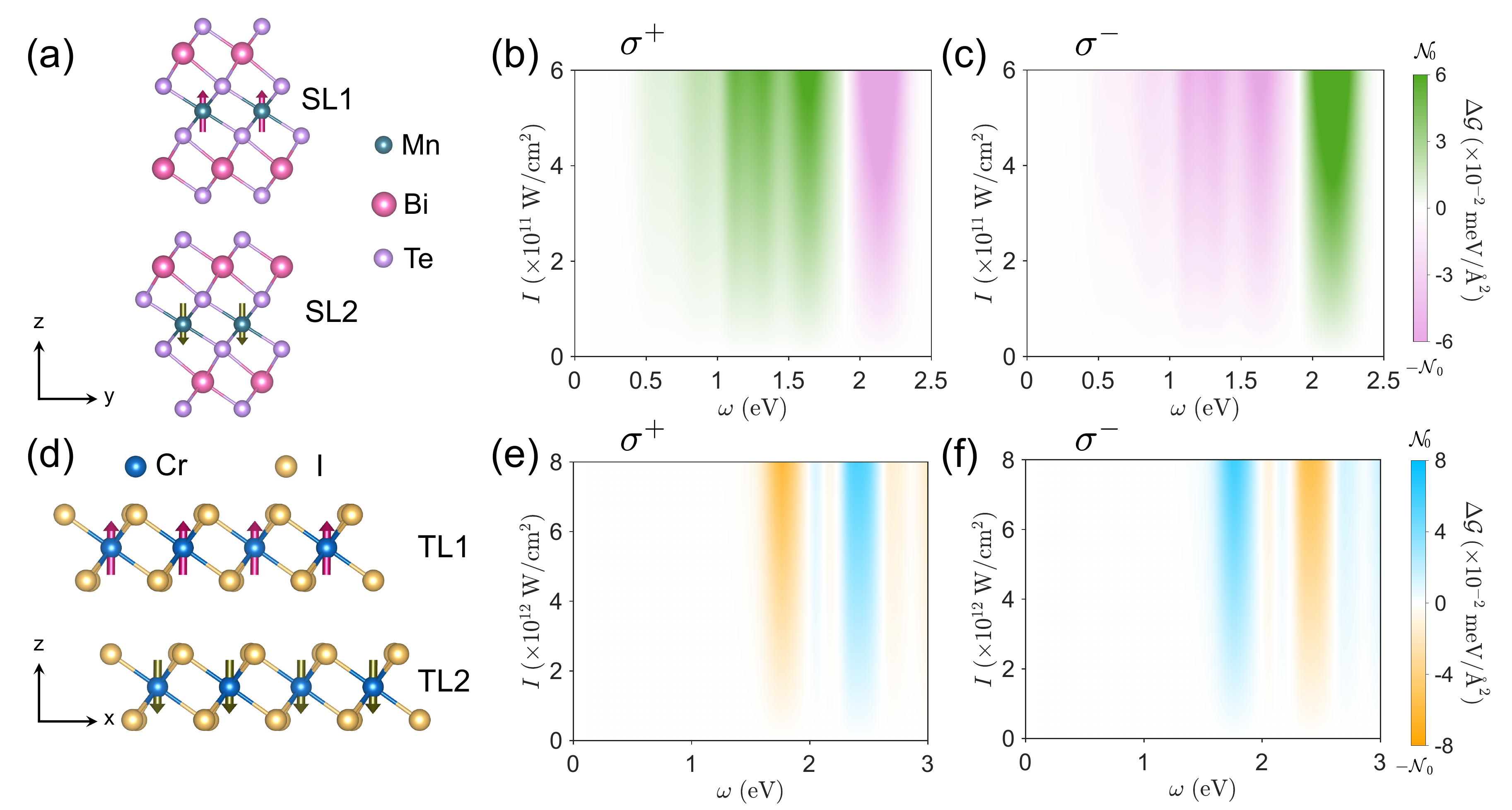}
    \caption{DFT results on realistic materials. Atomic geometry of (a) $\mathrm{MnBi}_2\mathrm{Te}_4$ bilayer, and the phase diagram under (b) $\sigma+$ and (c) $\sigma-$ light as functions of incident photon energy and light intensity. The light pulse duration is assumed to be $0.1\,\mathrm{ps}$. The color scheme represents free energy difference, with pink and green indicate $-\mathcal{N}_0$ and $\mathcal{N}_0$ is preferable under light. (d)-(f) The results for $\mathrm{CrI}_3$ bilayer under $\sigma+$ and $\sigma-$ light irradiation. The $\mathcal{N}_0$ configuration is plotted in (a) and (d). SL and TL stand for atomic septuple layer and trilayer, respectively.}
    \label{fig:dft}
\end{figure*}

\textit{Realistic materials.} We apply first-principle DFT calculations to adopt our theory into realistic materials, namely, $\mathrm{MnBi}_2\mathrm{Te}_4$ and $\mathrm{CrI}_3$. To reduce computational demands, we take bilayer systems here, which would give qualitatively the same results for thicker systems. The computational details can be found in SM \cite{supp}. As the conventional PBE functional \cite{Perdew08PRL,Perdew96PRL} underestimates the bandgap values, we adopt hybrid functional HSE06 \cite{Heyd03JCP,Heyd06JCP} corrections here. The atomic structure of $\mathrm{MnBi}_2\mathrm{Te}_4$ and its $\mathcal{N}_0$ configuration is plotted in Fig. \ref{fig:dft}(a). Our HSE06 calculated bandgap is $E_g=0.454\,\mathrm{eV}$, consistent with previous results \cite{Xue23NSR,Trang21ACSNano}. In the bilayer model, the energy difference can be expressed in a compact way
\begin{equation}\label{eq:bilayer}
    \begin{split}
        \Delta\mathcal{G}^{\sigma\pm}(\omega)=&\mathcal{G}_{-\mathcal{N}_0}^{\sigma\pm}(\omega)-\mathcal{G}_{\mathcal{N}_0}^{\sigma\pm}(\omega) \\
        =&\frac{d}{4c}I\left[\chi_{-\mathcal{N}_0}^{\sigma\pm,'}(\mathrm{L}_2,\omega)A_{-\mathcal{N}_0}^{\sigma\pm}(\mathrm{L}_1,\omega)\right. \\
        &\left.-\chi_{\mathcal{N}_0}^{\sigma\pm,'}(\mathrm{L}_2,\omega)A_{\mathcal{N}_0}^{\sigma\pm}(\mathrm{L}_1,\omega) \right].
    \end{split}
\end{equation}
Under $\mathcal{C}_{3z}$ and $\mathcal{C}_{2x}\mathcal{T}$, this can be furthermore simplified into competing between diagonal and nondiagonal multiplications of susceptibility and absorbance between the two layers (see Supplemental Material \cite{supp}).

The layer-resolved optical responses are shown in Fig. S4 \cite{supp}. Our DFT results are well-converged and in fully agreement with previously mentioned symmetry constraints. The top layer ($\mathrm{L}_1$) absorbs light on the order of $<\sim 5\%$, giving that attenuating factor $\lambda>\sim 95\%$. This would not generate enough heat in such a direct bandgap semiconductor, as most excited electron-hole pairs would recombine through direct radiative approaches. Hence, we ignore the thermal effect, which is in line with experimental observations \cite{Qiu23NM}. As seen in Eq. (\ref{eq:bilayer}), it is essential to lift the energy degeneracy between $\mathcal{N}_0$ and $-\mathcal{N}_0$ with nonzero attenuation. We then depict the energy difference between $\mathcal{N}_0$ and $-\mathcal{N}_0$, under $\sigma+$ and $\sigma-$ light [Figs. \ref{fig:dft}(b) and \ref{fig:dft}(c)]. Similar as in the model calculation, when the photon energy $\hbar\omega$ is below $E_g$, the two configurations are always degenerate, even though light scattering exists. When the light intensity $I$ reaches $5\times 10^{11}\,\mathrm{W}/\mathrm{cm}^2$ (corresponding to an electric field magnitude of $E_{\omega}=0.14\,\mathrm{V}/\angstrom$), the energy difference increases to be $0.06\,\mathrm{meV}/\angstrom^2$, which is about twice that of the magnetocrystalline anisotropy energy in bilayer $\mathrm{MnBi}_2\mathrm{Te}_4$ (calculated to be $0.03\,\mathrm{meV}/\angstrom^2$, consistent with previous works \cite{Otrokov19PRL}). Such a light intensity can be achieved in the pulse form experimentally. For thicker multilayers, such an intensity can be furthermore reduced. One also note that once the N\'eel vector flips, it would not reverse back, indicating a non-volatile phase transition.

It is interesting to note that incident light frequency (or wavelength) could change the N\'eel direction preference. According to our calculations, when the photon energy $\hbar\omega$ is below $1.9\,\mathrm{eV}$ (corresponding to wavelength $>650\,\mathrm{nm}$), $\sigma+$ light could induce the $\mathcal{N}_0$ state, while $\sigma-$ light gives the opposite $-\mathcal{N}_0$ configuration. If one increases the photon energy, the ground state reverses. At shorter wavelength ($<650\,\mathrm{nm}$), $\sigma+$ irradiation would yield $-\mathcal{N}_0$ and $\sigma-$ drives the $\mathcal{N}_0$ to be more stable. We briefly compare our results with a recent experimental work \cite{Qiu23NM}, in which Qiu \textit{et al.} have shown that $\sigma+$ light with wavelength shorter than $\sim 600-700\,\mathrm{nm}$ could result in the negative N\'eel vector state (probed by magnetic circular dichroism). At longer wavelength, the preferred N\'eel vector reverses to be upward. If the light handedness is changed, the N\'eel vector direction also varies. This clearly demonstrates that the photo-magnetic interactions dominate during such a process, while thermal effect can be eliminated. Note that the light handedness and transition frequency as predicted in our theory is fully consistent with these experimental facts. Our numerical calculations are free of any experimental or experienced parameters, suggesting the validity of this microscopic mechanism.

As this nonreciprocal light scattering theory is not limited in axion insulators, we then use another $A$-type AFM, bilayer $\mathrm{CrI}_3$, to illustrate our theory. This material has been demonstrated theoretically and experimentally to possess an out-of-plane $A$-type AFM configuration. Our HSE06 calculation gives $E_g=1.39\,\mathrm{eV}$. Hence, when $\hbar\omega\leq E_g$, the two N\'eel vectors remain to be degenerate. As the $E_g$ is larger, light response functions in $\mathrm{CrI}_3$ are in general smaller than $\mathrm{MnBi}_2\mathrm{Te}_4$. Hence, the light intensity required for flipping N\'eel vector is larger. Above the $E_g$, light attenuation arises, but the layer-resolved absorbance remains to be within $1\%$, with a large $\lambda$ value (Fig. S5 \cite{supp}). Above the bandgap, we see that $\sigma+$ light favors $-\mathcal{N}_0$, which can be converted to $\mathcal{N}_0$ under $\sigma-$ light irradiation. The energy difference could reach $0.08\,\mathrm{meV}/\angstrom^2$ under light of $I=8\times 10^{12}\,\mathrm{W}/\mathrm{cm}^2$ ($E_{\omega}=0.55\,\mathrm{V}/\angstrom$), which is twice of the magnetocrystalline anisotropy energy (calculated to be $0.04\,\mathrm{meV}/\angstrom^2$). Further increasing the photon energy to above a critical photon energy $\hbar\omega_{c1}=\sim 2.1\,\mathrm{eV}$ (wavelength of $590\,\mathrm{nm}$) reverses the light handedness and its preferred N\'eel vector, namely $\sigma+$ ($\sigma-$) gives $\mathcal{N}_0$ ($-\mathcal{N}_0$). Such a trend flips again near a second critical photon energy $\hbar\omega_{c2}=2.6\,\mathrm{eV}$. At higher $\hbar\omega$, $\Delta\mathcal{G}^{\sigma\pm}$ becomes smaller in magnitude, ascribed by the weaker nonreciprocal light scattering. Hence, we propose to use light with wavelength in between $~500-830\,\mathrm{nm}$ to control N\'eel vector direction in AFM-$\mathrm{CrI}_3$.

\textit{Discussion.} The simultaneous breaking of $\mathcal{P}$ and $\mathcal{T}$ here leads to optical diode effect, in analogous to semiconducting diode, and yields the nonreciprocal directional dichroism (NDD). Hence, this is another NDD mechanism in addition to the optical magnetoelectric (OME) coupling \cite{Kocsis18PRL,Sato20PRL,Kimura22NC}. In general, the OME could occur with low photon frequency, even at static limit, known as magetoelectric responses and closely related to toroidal moment $\bm T$ \cite{Ederer07PRB,Spaldin08JPCM}. 
We also note that layer-dependent susceptibility have been demonstrated to exhibit promising chiral and NDD magnetoelectric responses \cite{Stauber18PRL,Gao20PRL,Liang23PRL}. Our mechanism does not require the axion field to switch the N\'eel vector, and hence its material platform is not limited by axion insulators. We would like to remark that for the $C$-type or $G$-type AFMs, light incident direction may be switched to $[110]$ or $[111]$ directions, depending on how parallel spin planes are aligned. This is beyond the scope of current work, and will be discussed elsewhere.

\textit{Conclusion.} In summary, we propose a nonreciprocal scattering mechanism to lift the energy degeneracy between N\'eel up and down configurations in $\mathcal{PT}$-AFMs under CPL illumination. We show that the light scattering effect (akin to the optical tweezer or ponderomotive potential) that has been largely overlooked could tilt the degeneracy of bistable N\'eel polarization. This is not limited to axion insulators that require strong diagonal magnetoelectric coupling coefficient. We illustrate our theory using a minimum $\bm k\cdot\bm p$ model, and \textit{ab initio} DFT calculations in two realistic materials, giving consistent results as observed in very recent experiments. Our theory provides an alternative route to boosting optical diode effect.

\begin{acknowledgments}
The authors acknowledge valuable discussions with Yang Gao, Wenbin Li, Yandong Ma, Chengwang Niu, Hua Wang, Chunmei Zhang, and Hong Jian Zhao. This work is supported by the National Natural Science Foundation of China (NSFC) under Grant No. 12374065. The Hefei Advanced Computing Center is acknowledged where the calculations are performed.
\end{acknowledgments}

\providecommand{\noopsort}[1]{}\providecommand{\singleletter}[1]{#1}%

\end{document}